\documentclass[12pt,preprint]{aastex}

\shorttitle{Magnetic field in supernova remnant SN~1987A}
\shortauthors{E.G. Berezhko \& L.T. Ksenofontov}

\begin{document}

\title {Magnetic field in supernova remnant SN~1987A}

\author{E.G.~Berezhko\altaffilmark{1}
        L.T.~Ksenofontov\altaffilmark{1}
}
\altaffiltext{1}{Yu.G. Shafer Institute of Cosmophysical Research and Aeronomy,
                     31 Lenin Ave., 677980 Yakutsk, Russia}

\email{berezhko@ikfia.ysn.ru}

\begin{abstract} 
A nonlinear kinetic theory of cosmic ray (CR) acceleration in
supernova remnants is employed to investigate the properties of
the remnant 
SN~1987A. It is shown that a
large downstream magnetic field $B_d\approx 10$~mG is required
to fit the existing observational data.
Such a strong field together with the strong shock modification due to
CR backreaction provides the steep 
and concave radioemission
spectrum and
considerable synchrotron cooling of high energy electrons which diminish
their X-ray synchrotron flux below the observed Chandra flux
which has to be considered as an upper limit for nonthermal
X-ray emission. The expected $\gamma$-ray energy flux at TeV-energies at the current epoch 
is $2\times 10^{-13}$~erg/(cm$^2$s).
\end{abstract}


%________________________________________________________________

\section{Introduction}

Supernova (SN) 1987A occurred in the Large Magellanic Cloud.
It has been extensively studied in all wavelengths from radio to $\gamma$-ray.
The initial short outburst of radio emission \citep{turtle87} is attributed
to the synchrotron emission of electrons accelerated by the SN shock propagated
in the free wind of
presupernova star, which was the blue supergiant (BSG) 
\citep{chf87}. After about 3 years
radio emission was detected  again \citep{ss92, gae97}
as well as the monotonically increased X-ray emission \citep{gro94, has96}.
This second increase of emission is attributed to
the entrance of the outer SN shock into the thermalized BSG wind and then
in the H~II
region occupied by much more dense matter, consists of the 
swept up wind of red
supergiant (RSG) progenitor star \citep{chd95}.

To describe the observed properties of nonthermal emission of
SN~1987A we use here a nonlinear
kinetic theory. It couples the particle acceleration process
with the hydrodynamics of the thermal gas \citep{byk96, bv00}.
Therefore
in spherically symmetric approach it is able to quite definitely
predict the evolution of gas density, pressure, mass velocity,
together with the energy spectrum
and the spatial distribution of cosmic ray (CR) 
nuclei and electrons at any given
evolutionary epoch $t$, and the properties of the nonthermal radiation
produced in SNRs due to these accelerated CRs. The application of this theory
to individual SNRs \citep[see][for a review]{ber05} 
has provided the 
explanation of the observed SNR properties 
under the assumption of strong interior 
magnetic field. It was also shown that this strong field leads to the concentration of
the highest-energy electrons in a very thin shell just behind the shock.
Recent observations with the Chandra and XMM-Newton X-ray telescopes in space
make it possible to resolve such filamentary structures, which are the 
result of strong synchrotron losses of the emitting multi-TeV electrons in
strong magnetic fields downstream of the outer SN shock
\citep{ber05}.  

Application of the nonlinear theory to SN~1987A \citep{bk00}
explained the properties of the observed radioemission. It predicted also the
continuous increase of radioemission with roughly constant rate during the period
from 1999 to 2006 yr, which was confirmed by further measurements \citep{manch02}.
 
We analyze here radio data for more extended period of observation
\citep{mg01,manch02} together with the Chandra
measurements of X-ray emission \citep{park04} which 
provides a strong constraint on the amount of
the nonthermal synchrotron emission at the same energies 0.5-10~keV. It is
demonstrated that this constrain gives the evidence of strongly enhanced
(amplified) magnetic field inside the SNR, that in turn provides a better fit of
radio data than in our previous study.

\section{Physical parameters of SNR 1987A}
We use canonical values of stellar ejecta mass
$M_{ej}=10M_{\odot}$, distance $d=50$~kpc, 
hydrodynamic explosion energy $E_{sn}=1.5\times 10^{51}$~erg
\citep[e.g.][]{mccray93}.
During an initial period the shell material has a broad distribution in
velocity $v$. The fastest part of this ejecta distribution can be described by
a power law $dM_{ej}/dv\propto v^{2-k}$. We use
a value $k=8.6$ appropriate for SN~1987A \citep{mccray93}.
The interaction of the ejecta with the circumstellar medium 
(CSM) creates
a strong shock there which heats the thermal gas and accelerates particles
diffusively to a nonthermal CR component of comparable energy density. 

Strongly nonuniform CSM is a result of interaction of
progenitors winds: dense slow RSG wind  and subsequent fast and
diluted BSG wind. 
During initial period after the explosion 
the SN shock propagated in the free BSG wind and reached after about 
a day $\sim 1000$
the termination shock in BSG wind, situated at the radial distance
$R_T=3.1\times 10^{17}$~cm \citep{bk00}. 
After that the SN shock propagates in the
thermalized BSG wind of density $\rho_B$. Considerable rapid decrease of the SN
shock speed occurred during the days $1500-2500$ \citep{gae97} 
showed that SN shock entered the H~II region
of density $\rho_R\gg \rho_B$. Therefore we model
the CSM density distribution at distances $r>R_T$ in the form
\begin{equation}
\rho_0=\frac{\rho_{B}+\rho_{R}}{2}-
\frac{\rho_{B}-\rho_{R}}{2}
\tanh\frac{r-R_C}{l_{C}},
\end{equation}
where $R_C=6.1\times 10^{17}$~cm is the contact discontinuity between two winds,
$l_{C}=0.05R_C$ is the scale of the smooth transition between them.

The analysis of the rapidly increasing X-ray emission provides the evidence that
H~II region in turn is not uniform: at day $\sim 6200$ the SN shock begins to
interact with the dense inner ring. Its radial behavior can be represented
in the form 
\begin{equation}
\rho_R=\frac{\rho_{R1}+\rho_{R2}}{2}-
\frac{\rho_{R1}-\rho_{R2}}{2}
\tanh\frac{r-R_R}{l_R},
\end{equation}
where $R_R=6.8\times 10^{17}$~cm, $l_R=0.08R_R$, $\rho_{R2}\approx 20
\rho_{R1}$. Such a gas radial profile is close to what was extracted
from the above analysis \citep{park04}.
We adopt here for the gas number density
$N_g=\rho/m_p$ the following values:
$N_g^B=0.29$~cm$^{-3}$, $N_g^{R1}=280$~cm$^{-3}$
and $N_g^{R2}=4000$~cm$^{-3}$. Here $m_p$ is the mass of proton.
The value $N_g^{R1}=280$~cm$^{-3}$
which is by a factor
of 1.5 lower than was used in our previous study \citep{bk00},
provides a good compromise between the SN shock dynamics seen in radio and
in X-ray emissions (see below). Note, that the actual structure of the H~II
region, especially the dense inner ring, is essentially nonspherically
symmetric. Nevertheless since our spherically symmetric theory
reproduces the SN shock evolution $R_s(t)$ consistent with the existing data,
one can expect that it gives a good estimate for the production of CRs and their
subsequent nonthermal emission.

A rather high downstream magnetic field strength
$B_d\sim 1$~mG is
needed to reproduce the observed steep radio spectrum
\citep{bk00}. We believe
that the required strength of the magnetic field
have to be attributed to
nonlinear field amplification at the SN shock by CR
acceleration itself. According to plasma physical
considerations \citep{lb00,bell04}, the existing CSM magnetic field can
indeed be significantly amplified at a strong shock by CR
streaming instabilities.
In fact, for all the
thoroughly studied young SNRs, the ratio of magnetic field
energy density $B_0^2/8\pi$ in the upstream region of the shock precursor to
the CR pressure $P_c$ is about the same
\citep{vbk05}. Here $B_0=B_\mathrm{d}/\sigma$ is the far upstream field
presumably amplified by CRs of highest energy, $\sigma$ is the total shock compression ratio.
Within an error of about 50 percent we have
$ %\begin{equation}
B_0^2/(8\pi P_c) \approx 5\times 10^{-3}.
$ %\end{equation}
CR pressure in young SNRs has a typical value $P_c\approx 0.5 \rho_0 V_s^2$,
therefore we adopt here upstream magnetic field 
\begin{equation}
B_0=\sqrt{2\pi \times 10^{-3}\rho_0 V_s^2}.
\end{equation}
Since the process of magnetic field amplification is 
not included in our theory
we simply postulate the existence of far upstream field $B_0$ given by Eq.(3).
The spectrum of CRs produced by strong
modified shock is very hard so that CRs with highest energies have a largest
contribution in their energy content. These the most energetic CRs produce
field amplification on their spacial scale that is the precursor size.
Therefore CRs with lower energies already 'see' the amplified field $B_0$. 
We will argue that such a high field,
that is by a factor of ten larger than used in our previous study \citep{bk00},
is indeed required to fit the X-ray data. It also gives a better fit of radio
data.

We start our consideration from the SNR evolutionary epoch $t=1000$~d, when the
outer SN shock has a radius $R_i=R_T$ and speed $V_i=28000$~km/s. These values
of $R_s$ and $V_s$ according
to our calculations \citep{bk00} correspond to the end of SN
shock propagation in the free BSG wind region $r<R_T$. We neglect the
contribution of CRs accelerated in the region $r<R_T$, because due to a high gas
density the number of CRs produced in the region $r>R_T$ very soon becomes
dominant.

\section{Results and Discussion}

\begin{figure}
\epsscale{0.5}
\plotone{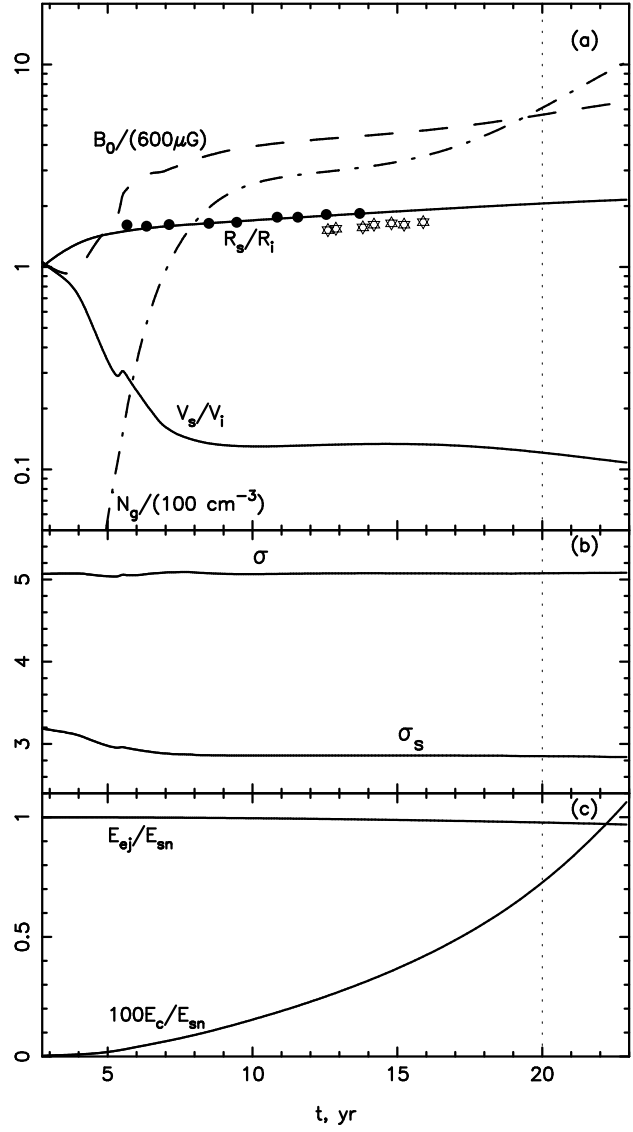}
\caption{(a) Shock radius $R_\mathrm{s}$, 
shock speed $V_\mathrm{s}$, 
gas density $N_g$ and upstream magnetic field $B_0$ at the current shock
position; (b) total
shock ($\sigma$) and subshock ($\sigma_\mathrm{s}$) compression ratios;
(c) kinetic energy of ejecta $E_{ej}$ and accelerated CR energy content $E_c$
as functions of time. 
The {\it dotted
vertical line} marks the current epoch. The observed radius of
the SN shock, as determined by radio \citep{manch02} 
and X-ray measurements \citep{park04}, are shown by circles and stars
respectively.
The scaling values are $R_i=R_T=3.1\times 10^{17}$~cm and $V_i=28000$~km/s
\label{f1}}
\end{figure}

Calculated shock radius $R_s$ and speed 
$V_s$ shown in Fig.\ref{f1}a as a function of
time are in satisfactory agreement with the values obtained on the basis of
radio and X-ray measurements. Note that radio data 
compared with X-ray data gives larger shock size
at any given time $t$ and calculated radius 
$R_s(t)$ goes between these two sets

To fit the spectral shape of the observed
radio emission we assume a proton injection rate
$\eta=3\times 10^{-3}$, which is a fraction of gas particles involved into the
acceleration at the from of SN shock. This leads to a significant nonlinear modification of
the shock: as it is seen in Fig.\ref{f1}b  total shock compression ratio
$\sigma\approx 5.3$ is essentially larger
and a subshock compression ratio $\sigma_s\approx 2.8$ is lower 
than classical value 4.

Since the SN~1987A is very young CRs despite of very efficient acceleration
accumulated only 7\% of the explosion energy $E_{sn}$ (see Fig.\ref{f1}c). 
Note that the injection is expected to be strongly suppressed at the 
quasiperpendicular part of the
shock, therefore
one should renormalize the results for the nucleonic spectrum, 
calculated within
the spherically symmetric model. The lack of symmetry in the actual 
SNR can be approximately
taken into account by a renormalization factor 
$f_{re}\approx 0.2$, which diminishes the nucleonic CR production efficiency, 
calculated in the spherical
model, and all effects associated with it \citep[see][for details]{vbk03}.
With  this renormalization the CRs inside SN~1987A
already contain
\begin{equation}
E_c=0.07f_{re}E_{sn}\approx 2\times 10^{49}~\mbox{erg}.
\end{equation}

Since during the last 15~yrs the shock speed 
and the gas density are almost constant CR
energy content grows roughly linearly with time. It gives the natural
explanation of the linear increase of the radio emission detected during this
evolutionary period.

\begin{figure}
\epsscale{1.0}
\plotone{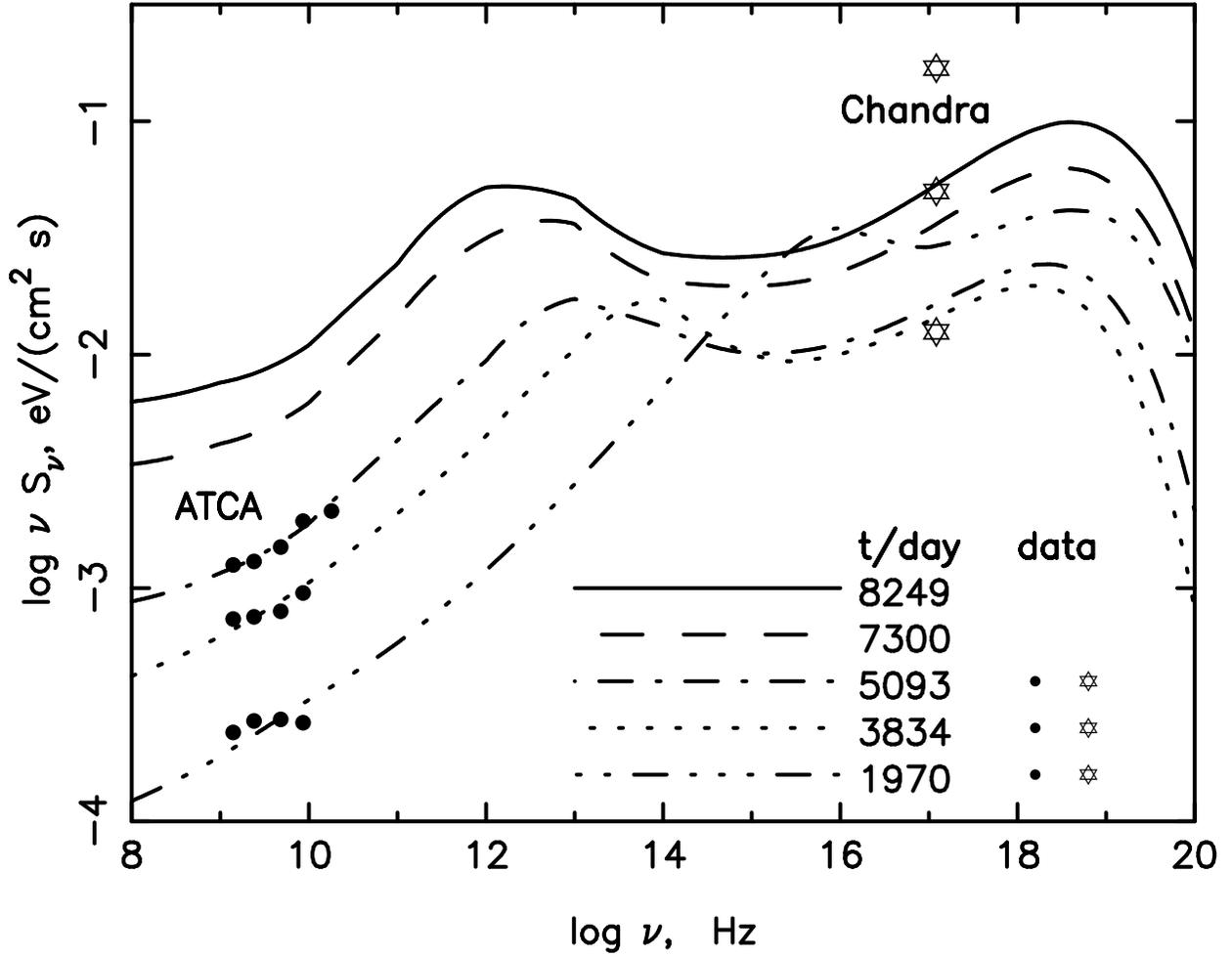}
\caption{Synchrotron energy spectrum of SN~1987A, calculated for 
the five evolutionary epochs. The 
{\it ATCA} radio \citep{manch02, mg01}
and Chandra X-ray \citep{park04} data for three epochs are shown as
well. Higher measured fluxes correspond to later epoch.
\label{f2}}
\end{figure}

Strongly modified SN shock generates CR spectrum 
$N\propto p^{-\gamma}$, which is very steep at momenta
$p<m_pc$, with index $\gamma =(\sigma_s+2)/(\sigma_s-1)\approx 2.7$.
CR electrons with such a spectrum produces synchrotron radioemission
spectrum $S_{\nu}\propto \nu^{-\alpha}$ with spectral index $\alpha
=(\gamma-1)/2\approx 0.9$, that very well corresponds to the experiment,
as it is seen in Fig.\ref{f2}, where we present synchrotron energy spectra
$\nu S_{\nu}$,
calculated for five subsequent epoch together with the experimental data.
Note that CR spectrum has a concave shape: it becomes flatter at higher momenta
$p$.
As a consequence synchrotron spectrum $S_{\nu}(\nu)$ is also concave as it is
clearly seen in Fig.\ref{f2} at $\nu<10^{12}$~Hz. Radio data reveal this property in
good consistency with theoretical prediction.

Strong downstream magnetic field $B_d\approx 15$~mG, that corresponds to the
upstream field $B_0\approx 3$~mG (see Fig.\ref{f1}), provides synchrotron cooling
of electrons with momenta $p>10m_pc$ that in turn makes 
synchrotron spectrum at high frequencies $\nu>10^{12}$~Hz
very steep (see Fig.\ref{f2}). 
Concave shape of electrons continuously produced at the shock front together with
their synchrotron cooling lead to a formation of two peaks in synchrotron energy
spectrum $\nu S_{\nu}$. The first one at $\nu \approx 10^{12}$~Hz corresponds to
CR electron momentum $p\approx 10m_pc$ above which synchrotron energy looses
are relevant, whereas the second peak at $\nu \approx 10^{18}$~Hz corresponds
to the maximum momentum $p\approx 10^4m_pc$ of accelerated electrons.
Under this condition calculated
synchrotron flux at frequency $\nu\approx 10^{17}$~Hz, which
corresponds to the photon energy $\epsilon_{\gamma}=0.5$~keV, 
is below the measured
flux at the epochs $t>3000$~d.
Since the contribution of the nonthermal radiation in the
observed X-ray emission of SN~1987A is not 
very well known \citep[e.g.][]{mich02},
the observed X-ray flux has to be considered as 
the upper limit for the expected nonthermal emission. 
At early epoch $t<2500$~d however the calculated flux
exceeds the measured one (see the curve, corresponding $t=1970$~d in
Fig.\ref{f2}).
This can be considered as indication, that the actual magnetic field
$B_0$ is few times larger then given by the Eq.(3).

\begin{figure}
\plotone{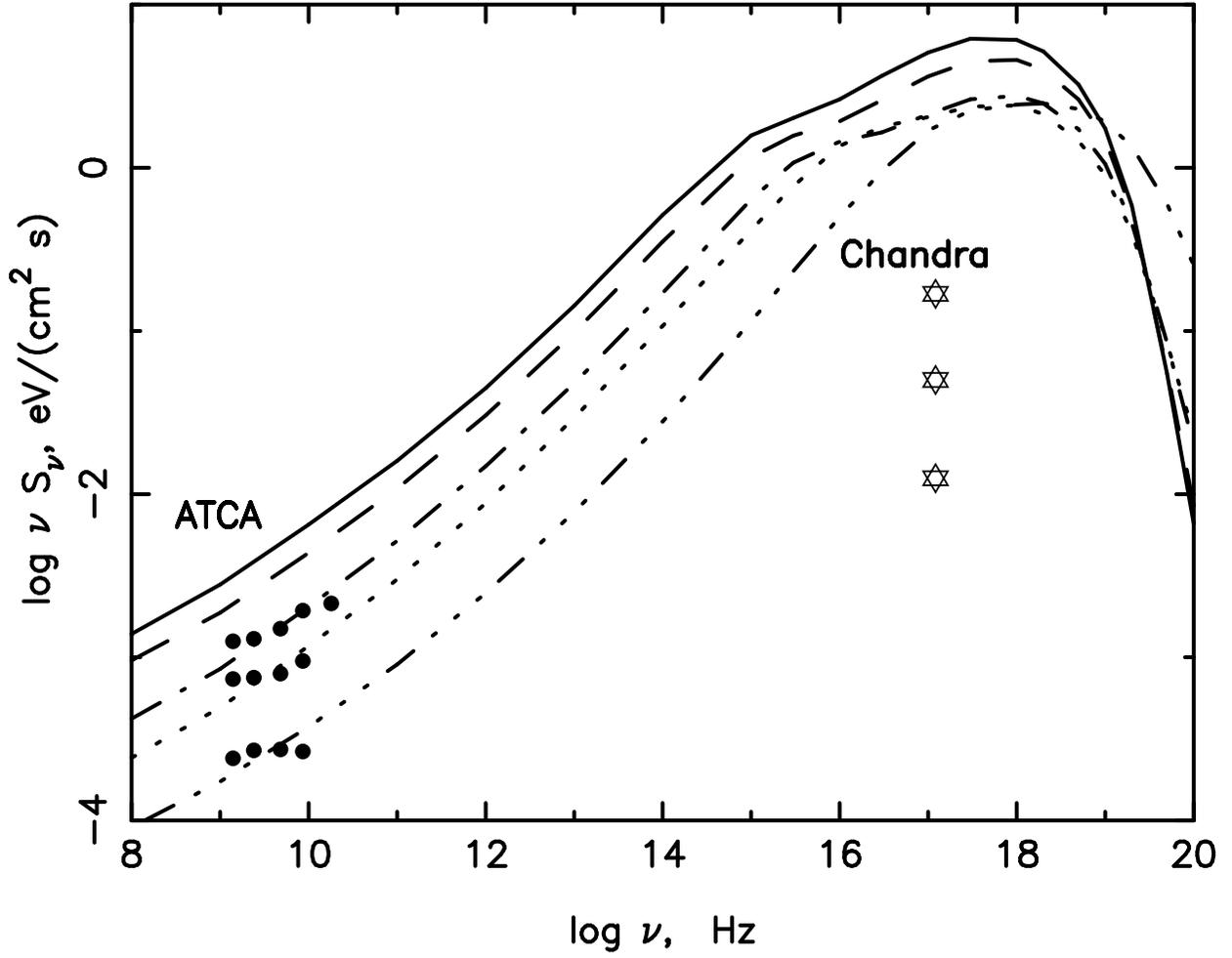}
\caption{Same as in Fig.\ref{f2}, but the calculations are for the 
case of constant upstream magnetic field $B_0=200$~$\mu$G.
\label{f3}}
\end{figure}

To illustrate the situation expected at considerably lower magnetic field we
present in Fig.\ref{f3} synchrotron energy spectra calculated at the same set of
parameters as before except magnetic field, which was taken $B_0=200$~$\mu$G
independent of time. Since the interior magnetic field $B_d \approx 2$~mG 
($\sigma \approx 10$) is
essentially lower in this case, synchrotron losses of high energy CR electrons
are considerably smaller compared with the previous case. Due to this fact
synchrotron spectra considerably exceeds at any given epoch the measured Chandra
flux. Therefore we can conclude, that the actual interior magnetic field
strength is not lower than $5$~mG.

Projected radial profile of the nonthermal X-ray emission at high
interior magnetic field $B_d=10$~mG is characterized by an extremely 
sharp peak of thickness
$L\approx 10^{-3}R_s$ just behind the SN shock. Since the thermal X-rays have much
wider radial distribution it provides the possibility 
for experimental determination
of the nonthermal X-ray contribution in the observed emission.
Experimental determination of the X-ray spectrum at $\nu =10^{18}-10^{19}$~Hz
($\epsilon_{\gamma}=5-50$~keV) is the another possibility to discriminate the
nonthermal emission. Such kind of measurements can be done with space
instrument Suzaku.

Completely different 
possibility to have the synchrotron flux, which fit the radio data and
goes below Chandra data, is low field scenario, when the
magnetic field is
as low as $B_0<2$~$\mu$G. In such a case the  
cutoff frequency of synchrotron
spectrum $\nu_{max}$ is lower than $10^{17}$~Hz and 
calculated fluxes go down exponentially at $\nu>\nu_{max}$
below Chandra points. However the value $B_0<2$~$\mu$G is unrealistically small
for such a dense medium, which we have in SN~1987A. 
In addition the effective energy of electrons
\begin{equation} 
\epsilon_e\approx 
5\sqrt{(\nu/\mbox{1~GHz})/(B_d/10~\mu\mbox{G})}\mbox{~~GeV}, 
\end{equation}
which radiate at $\nu \sim 1$~GHz in the interior field $B_d=10$~$\mu$G
is about 5~GeV. The electron spectrum at GeV-energies
is characterized by a power low index $\gamma\approx 2$ that is considerably
smaller than what is required for the observed radio spectrum.
Note that as it is seen in Fig.\ref{f3} even the field $B_d\approx 2$~mG is too small
to have a good fit of radio data as in Fig.\ref{f2}.
Therefore low field scenario should be rejected.

\begin{figure}
\plotone{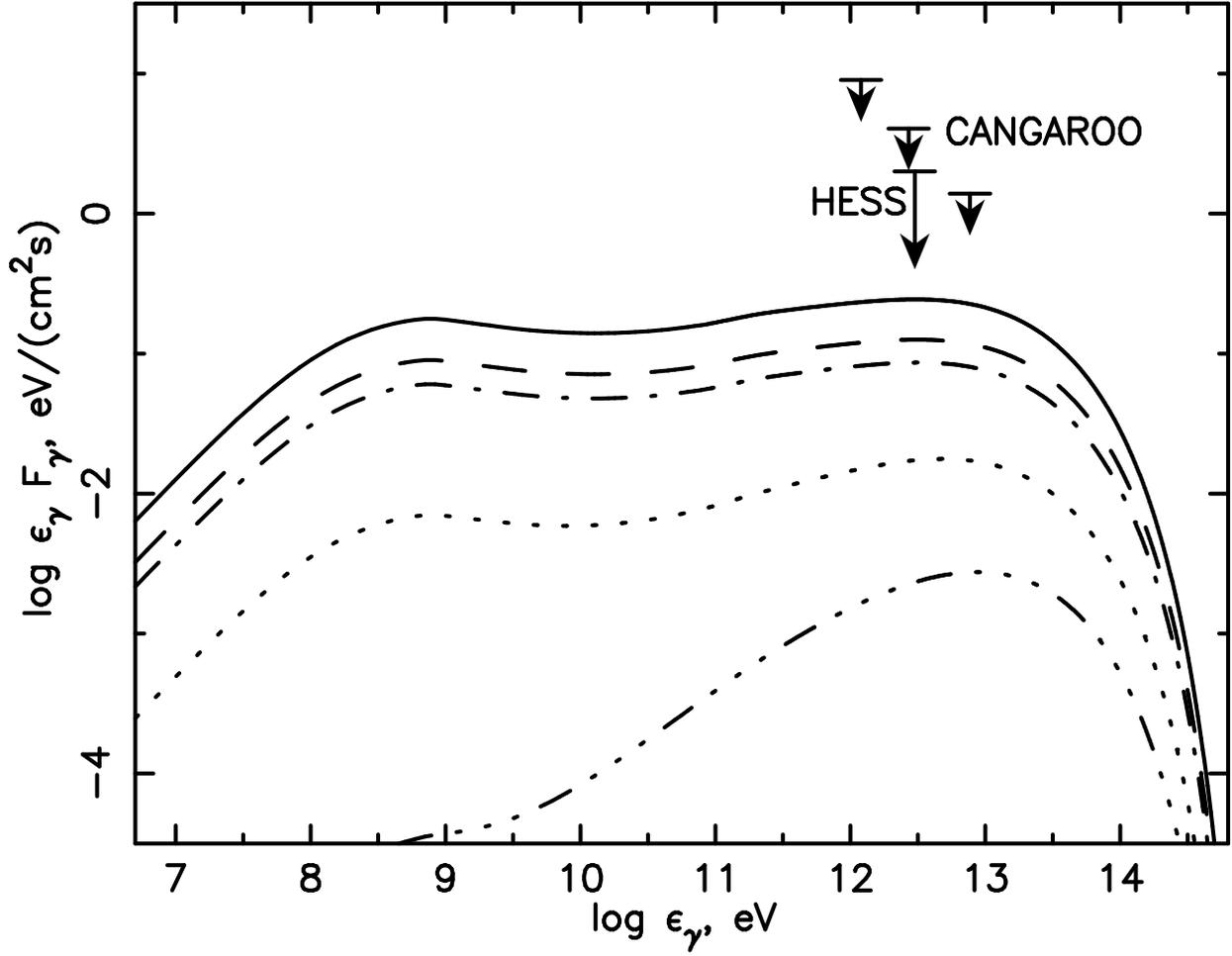}
\caption{Integral $\gamma$-ray energy flux from SN~1987A, calculated for the same five
epoch as in Fig.\ref{f2}. CANGAROO \citep{eno03} and HESS \citep{rowell04}
upper limits are shown as well.
\label{f4}}
\end{figure}

Calculated $\gamma$-ray integral flux shown in Fig.\ref{f4} 
at all energies is dominated by the $\pi^0$-decay component.
Since the SN shock is strongly modified $\gamma$-ray spectrum at
energies $\epsilon_{\gamma}>0.1$~TeV is very hard: $F_{\gamma}\propto
\epsilon_{\gamma}^{-0.9}$. 
At the current epoch the expected $\gamma$-ray energy flux at
TeV-energies is about 
$\epsilon_{\gamma}F_{\gamma}\approx 2\times 10^{-13}$~erg/(cm$^2$s) 
and during the next four years it
expect to grow by a factor of two. This flux is by a factor of five
lower compared with our previous prediction \citep{bk00}, because we use here
renormalization factor $f_{re} =0.2$, which reduces the number of accelerated
CRs calculated within spherically symmetric approach. 
The existence of strongly asymmetric CSM structure, which is dense inner ring,
makes our prediction of $\gamma$-ray flux uncertain. According to the rough estimate this
uncertainty is not very large, about a factor of two, due to the stronger
SN shock deceleration in denser medium.
At the moment there are only upper limits of TeV emission obtained by CANGAROO
\citep{eno03} and HESS \citep{rowell04} instruments (see Fig.\ref{f4}).

\section{Summary}
The kinetic nonlinear model for CR acceleration in SNRs has been in
detail applied to SN~1987A, 
in order to compare its results with 
observational properties.We find that quite a reasonable 
consistency with most of the observational
data can be achieved. 

The evidence of the
efficient CR production leading to a strong shock modification 
comes from radio data. Significant shock modification
leads to the steep and concave CR spectrum 
which very well fits the observed radioemission spectrum, under
the condition of extremely high downstream magnetic field strength 
$B_d\sim 1$~mG. To be consistent also with the Chandra measurements
of the X-ray flux, which has to be
considered as an upper limit for the nonthermal X-ray emission, even larger
interior field $B_d\approx 10$~mG is needed. Such a high field provides
a strong synchrotron losses of CR electrons emitting nonthermal
X-rays, that makes the high frequency part of the synchrotron spectrum much
more steeper and provides a consistency with the experiment.

The expected $\pi^0$-decay $\gamma$-ray energy flux 
at the current epoch is about
$\epsilon_{\gamma}F_{\gamma}\approx 2\times 10^{-13}$
~erg/(cm$^2$~s) at energies
$\epsilon_{\gamma}=0.1-10$~TeV. Therefore the detection of
$\gamma$-ray emission at these energies would imply clear evidence 
for a hadronic origin and for a strong magnetic field amplification inside
SN~1987A.

\clearpage

\clearpage

\clearpage

\clearpage

\end{document}